\documentclass[11pt,a4paper]{article}
\pdfoutput=1
\usepackage{jheppub}
\usepackage{epsfig}
\numberwithin{equation}{section}
\usepackage{amssymb}
\usepackage{subfigure}

\usepackage{feynmp}

\def \abs#1{\left\vert#1\right\vert}
\newcommand{\be}{\begin{equation}}
\newcommand{\ee}{\end{equation}}
\newcommand{\bea}{\begin{eqnarray}}
\newcommand{\eea}{\end{eqnarray}}
\newcommand{\bwt}{\begin{widetext}}
\newcommand{\ewt}{\end{widetext}}

\title{Vector-like Fields, Messenger Mixing and the Higgs mass in Gauge Mediation}

\author[a]{Willy Fischler}
\author[a]{and Walter Tangarife}
\affiliation[a]{Department of Physics and Texas Cosmology Center\\ The University of Texas at Austin,
TX 78712.}
\emailAdd{fischler@physics.utexas.edu}
\emailAdd{wtang@physics.utexas.edu}

\abstract{ In order to generate, in the context of gauge mediation, a Higgs mass around 126 GeV  that avoids the little hierarchy problem,  we explore a set of models where the messengers are directly coupled to new vector-like fields at the TeV scale in addition to the usual low energy degrees of freedom. We find that in this context, stop masses lighter than 2 TeV and large $A$-terms are generated, thereby improving issues of fine tuning.}


\preprint{UTTG-26-13, TCC-022-13}
\begin{document}
\maketitle

%%%%%%%%%%%%%%%%%%%%%%%%%%%%%%%%%%%%%%%%%%%%%%%%%%%%%%%%%%%%%%%%%%%%%%%%%%%%

%          Table of contents automatic !!!                                 %

%%%%%%%%%%%%%%%%%%%%%%%%%%%%%%%%%%%%%%%%%%%%%%%%%%%%%%%%%%%%%%%%%%%%%%%%%%%%

\section{\bf Introduction}

The last two years have been remarkably exciting for both experimentalists and theorists working in high energy physics due to the discovery of the Higgs particle \cite{Aad:2012tfa,Chatrchyan:2012ufa}. However, a Higgs mass of $126\,{\rm GeV}$ poses intriguing questions, for theorists, about the naturalness of minimal supersymmetric models  \cite{Baer:2011ab,Hall:2011aa,Arvanitaki:2013yja}. Due to the fact that, at tree level, the mass of the Higgs is bounded by the $Z$ boson mass, a large contribution to $m_h$ must come from radiative corrections, which are dominated by the stop fields. Thus, it would require to have very heavy stop masses in order to generate such mass value. However, large stop masses make the soft parameter $-m_{H_u}^2$ large as well, which is straining the electroweak symmetry breaking condition $$m_Z^2\approx -2(m_{H_u}^2 +\mu^2),$$ implying a large amount of fine tuning to achieve the proper cancellation between $-m_{H_u}^2$ and $\mu^2$\footnote{To gain insight on the measurement of fine tuning, see \cite{Barbieri:1987fn,deCarlos:1993yy,Chankowski:1997zh,Kitano:2005wc} and references in \cite{Hall:2011aa}.}. This is known in the literature as  the ``little hierarchy problem".
 
There have been several approaches to raising the mass of the Higgs in the literature. In one strategy, keeping SUSY minimal, the use of large trilinear $A-$terms has been studied as a way to avoid the requirement of very heavy stops. This is, however, somewhat difficult to achieve in the standard gauge mediation of supersymmetry breaking scenario (GMSB) \cite{Dine:1981gu,Nappi:1982hm,Dine:1982zb,AlvarezGaume:1981wy}, and would require very high messenger masses or a modification of the mediation mechanism. For instance, GMSB has been modified in models where mixing between the low energy degrees of freedom and the messenger fields is proposed proposed as a way to generate non-zero $A-$terms at the messenger scale \cite{Chacko:2001km,Komargodski:2008ax,Albaid:2012qk,Kang:2012ra,Grajek:2013ola,Byakti:2013ti,Craig:2013wga,Evans:2013kxa,Abdullah:2012tq,Kang:2012sy} \footnote{In \cite{Krauss:2013jva}, the Higgs mass is raised by modifying the gauge symmetry at the messenger scale.}.

In a different approach, extensions of the minimal supersymmetric standard model (MSSM) have been introduced in order to alleviate the little hierarchy problem. A clear example of such attempts is the next-to-MSSM (NMSSM, see \cite{Ellwanger:2009dp} for a review). Another proposal that falls in this category is the addition of a set of vector-like fields that couple to the Higgs multiplets. This raises the mass of the Higgs without requiring very heavy superpartners \cite{Babu:2004xg,Babu:2008ge, Martin:2009bg, Graham:2009gy, Martin:2010dc,Endo:2011mc,Endo:2011xq,Martin:2012dg,Yokozaki:2013eca,Chang:2013eia}. This type of extensions of the low energy matter content provides a spectrum with low masses that can be tested in the near future at the LHC.
 
In this work, we explore a model that combines these two approaches. Minimal GMSB is modified to include Yukawa couplings between the MSSM and the messenger sector and, at the same time, there are additional vector-like fields at the TeV scale. A similar approach was used several decades ago by Dine and Fischler \cite{Dine:1983xx} to generate electroweak symmetry breaking; however in their approach, the vector matter was much heavier. In our case, we provide a microscopic completion for a type of models presented in \cite{Martin:2009bg} in the context of gauge mediated supersymmetry breaking, with the additional feature of having large $A-$terms in the effective low energy Lagrangian, which leads to a Higgs mass consistent with the observed values at the LHC. We explore the parameter space and find a set of regions that yield a 126 GeV Higgs mass with stop masses under 2 TeV. We argue that the fine tune problem is substantially improved in this kind of scenario, compared to minimal GMSB, as there is no need for large stop masses and $m_{H_u}^2$ gets a smaller radiative contribution from stops and vector-like sfermions.

This article is structured as follows: In section \ref{sec:model}, we introduce the model with the new fields. We also present the soft terms calculated at the messenger scale. In section \ref{higgsmass}  we show the correction the the Higgs mass due to the new fields and the results obtained in the numerical analysis. We close with some conclusions and we include two appendices with the RGEs of the model and the complete expressions for the soft masses.

%%%%%%%%%%%%%%%%%%%%%%%%%%%%%%%%%%%%%%%%%%%

%%%%%%%           MODEL SECTION              %%%%%%%%%%%%

%%%%%%%%%%%%%%%%%%%%%%%%%%%%%%%%%%%%%%%%%%%

\section{The model}\label{sec:model}

In this paper, we introduce a set of new vector-like chiral superfields that are charged under the Standard Model (SM) gauge symmetry group. This is motivated by the work presented in \cite{Dine:1983xx}. There, new superheavy superfields were added in order to achieve the breaking of the gauge symmetry $SU(2)\times U(1)$ by generating negative values of $m_{H_u}^2$ in the Higgs potential. In the present case, we consider similar new superfields with  vector-like masses somewhere between 500 GeV and 1.4 TeV. The addition of these new superfields is expected to lift the mass of the Higgs through radiative corrections in such a way that very large stop masses are unnecessary. Furthermore, as done in \cite{Dine:1983xx},  we allow the MSSM fields and the new vector-like matter to interact directly with the messenger sector through Yukawa couplings. This mixing with the messengers will generate large trilinear terms ($A-$terms) that contribute to the Higgs mass enhancement.  

The new vector-like superfields can be arranged in complete representations of $SU(5)$. Here, we choose to use a pair of 10 dimensional representations $\mathbf{10} \,+\,\overline{\mathbf{10}}$,
\bea 
\mathbf{10}  &=& \Phi(3,2)_{1/6} + \Psi(\bar{3},1)_{-2/3}+ \chi(1,1)_1 \label{10-reps}\\ 
\mathbf{\overline{10}}  &=& \overline{\Phi}(\bar{3},2)_{-1/6} + \overline{\Psi}(3,1)_{2/3}+ \overline{\chi}(1,1)_{-1}. \nonumber 
\eea 
On the other hand, we use a pair of $\overline{\mathbf{5}} \,+\,\mathbf{5}$ $SU(5)$ representations in the messenger sector,
\bea
\mathbf{5}  &=& A(3,1)_{-1/3} + B(1,2)_{1/2}. 
\eea 
These messengers couple to the spurion $X$, which generates the messenger mass and breaks SUSY through its expectation value, $\langle X \rangle = M+\theta^2 F$, in the superpotential 
\be 
W_{\rm X} = X(\lambda_A A\overline{A} + \lambda_B B\overline{B}) \label{WX}.
\ee 

We consider a superpotential that connects three sectors: the MSSM\footnote{In this work we use the approximation in which only the third generation of the MSSM is included, and the only relevant mixing with the messengers occurs for the third generation and the new vector-like fields. This can be justified by using a flavor $U(1)$ symmetry as used in \cite{Albaid:2012qk}. Also, \cite{Chacko:2001km} describes a construction where the relevant messenger-matter couplings are those involving the third family as the result of integrating out an extra dimension.}, the new vector-like superfields, and the messengers fields. 
\bea
W &=& W_{\rm MSSM} \,+\, M_{10}(  \Phi\overline{\Phi}+\Psi \overline{\Psi}+ \chi \overline{\chi})  + h_1 \Psi H_u \Phi - h_2 \overline{\Psi} H_d  \overline{\Phi}\\
 & &   + \lambda_{1,B} \Psi B \Phi + \lambda_{2,B} \overline{\Psi} \overline{B} \overline{\Phi}  + \kappa_{1,B} Q U B + \kappa_{2,D} Q D \overline{B} +\kappa_{2, U} U D \overline{A}  \nonumber \\
 && +\lambda_{3,\Phi} \Phi \overline{A} \overline{B}   +\lambda_{4,\Phi} \overline{\Phi} A B + \lambda_{\chi} \Psi A \chi + \lambda_{\overline{\chi}} \overline{\Psi}\overline{ A} \overline{\chi}
 \,+\,W_{\rm X}, \nonumber \label{SuperPotential}
\eea
where the $SU(3)_C$ and $SU(2)_L$ indices have been contracted in the usual way, i.e. $\Phi \overline{\Phi} \equiv \epsilon^{\alpha \beta} \Phi^{a \alpha} \overline{\Phi}_a^\beta $, with $\alpha, \beta =1,2$ and $a=1,2,3$. The negative sign in front of $h_2$ is not necessary; however, it facilitates our analysis in analogy to $y_b$ in the MSSM. We assign $R-$parity $P_R=+1$ to the new vector-like fields to prevent the low energy vector-like fields from mixing with the MSSM quarks or leptons in the superpotential. It is worth mentioning that, in this work, we are allowing all the couplings in Equation (2.4) to be non-zero; this is different from previous works, where only one non-zero coupling was consider at a time.

\subsection{Effective mass terms}

Supersymmetry is broken due to the non-zero  F-term $\langle F_X\rangle= F$. This generates soft mases for the MSSM fields as well as for the new vector-like fields. Besides the soft masses generated through the usual gauge mediation mechanism (GMSB), there is an additional contribution to the masses of the sfermions due to the Yukawa couplings to the messenger fields. The calculation of these soft masses follow the same methodology presented in \cite{Evans:2013kxa}. 

This modified gauge mediation mechanism results in the soft Lagrangian
\begin{align}
-\mathcal{L}_{\rm soft} &= - \mathcal{L}_{\rm MSSM,soft} + m_{\tilde{\Phi}}^2 |\Phi |^2 + m_{\tilde{\overline{\Phi}}}^2 |\overline{\Phi} |^2+m_{\tilde{\Psi}}^2 |\Psi |^2 + m_{\tilde{\overline{\Psi}}}^2 |\overline{\Psi} |^2+ m_{\tilde{\chi}}^2 |\chi |^2 + m_{\tilde{\overline{\chi}}}^2 |\overline{\chi} |^2 \nonumber \\
  & + \left( b_\Phi \Phi \overline{\Phi} +b_\Psi \Psi \overline{\Psi} + a_1 \Psi \Phi H_u   -a_2 \overline{\Psi} \overline{\Phi} H_d   + h.c. \right), \label{SoftL}
\end{align}
where $a_i\equiv h_i A_i$, for $i=\,1,\,2,\,t,\,b,\,\tau$. These soft parameters are calculated at the messenger scale, $M$, and evolved down to the electroweak scale through the running of the renormalization group. For simplicity, we write here just the leading contributions to the soft masses of the vector-like field $\Phi$ coming from the gauge and Yukawa interactions with the messengers. In Appendix \ref{appendix1}, we present a general formula to compute these soft terms. Also, for simplicity, we make $\lambda_A = \lambda_B =1$.

\begin{align}
m^2_{\tilde{\Phi},\, {\rm gauge}}=& \frac{1}{8\pi^2} \left(\frac{F }{M} \right)^2\left[ \frac{4}{3} \alpha_3^2 +\frac{3}{4} \alpha_2^2 +\frac{3}{5} Y_{\Phi}^2 \cdot \frac{6}{5} \left(\frac{1}{9} +\frac{1}{4} \right) \alpha_1^2  \right] ,\\
m^2_{\tilde{\Phi},\, {\rm Yuk}} =& \frac{1}{256\pi^4} \left(\frac{F }{M} \right)^2 \left[ 6\lambda_{3,\Phi}^4+\lambda_{3,\Phi}^2\lambda_{3,\chi}^2 +  3\lambda_{1,B}( \kappa_{1,B}^2+h_1^2)  + 6\lambda_{1,B}^4\right.\\ 
  & \left.  + 6 \kappa_{1,B}\, y_t\, h_1 \,\lambda_{1,B} +(4\lambda_{3,\Phi}^2-2h_1^2)\lambda_{3,\Psi}^2  - \frac{4\pi}{3}\alpha_1(13 \lambda_{1,B}^2-56 \lambda_{3,\Phi}^2)  \right. \nonumber \\
  & \left.   -6\pi \alpha_{2} (\lambda_{1,B}^2+8\lambda_{3,\Phi}^2)-\frac{32\pi}{3}\alpha_{3}(\lambda_{1,B}^2+8\lambda_{3,\Phi}^2) \right]. \nonumber 
\end{align}

At the messenger scale, the $A-$terms are non-zero due to the Yukawa couplings to the messengers. For instance, $A_1$ takes the value
\begin{equation} 
A_{1} = -\frac{1}{16\pi^2}\left(\frac{F }{M} \right) (3\lambda_{1,B}^2+ \lambda_{3,\Phi}^2). 
\end{equation}  This implies that, at low energies, there is a significant mixing between the scalar superpartners. 

% new fields that contribute to the mass of the Higgs at one-loop level as will be discussed in section \ref{higgsmass}. 

After calculating the soft masses and $A-$terms, we run down the renormalization group from the messenger scale to the low energies where electroweak symmetry breaking is computed and the Higgs mass is obtained, as shown in the next section. The RGEs corresponding to this model are presented in the appendix \label{appendix2}.

In the effective theory, we have a set of electrically neutral fermionic fields that couple to the neutral components of the Higgs multiplets and with mass matrix  
\be 
m_F^2 = 
\begin{pmatrix}
\mathcal{M_F}\mathcal{M_F}^\dagger   &  \\
  &  \mathcal{M_F}^\dagger \mathcal{M_F} \\
  \end{pmatrix},\,\,{\rm with} \,\,
\mathcal{M_F} = 
\begin{pmatrix}
M_{10}  & h_1 H_u^0 \\
h_2 H_d^0  & M_{10} \\
  \end{pmatrix}. \label{mf}  
 \ee
On the other hand,  the mass matrix for the electrically neutral scalars coupled to the neutral Higgs bosons is given by 
\be 
m_S^2 = m_F^2 +  
\begin{pmatrix}
m^2_\Phi + \Delta_\Phi  & 0 & b^*_\Phi & a_1^* v_u - h_1 \mu v_d  \\
 0 &  m^2_{\Psi}+\Delta_\Psi  &  a_2^* v_d - h_2 \mu v_u & b_\Psi^*  \\
b_\Phi &  a_2 v_d - h_2 \mu^* v_u  &  m_{\bar{\Phi}}^2+ \Delta_{\bar{\Phi}}   & 0  \\
 a_1 v_u - h_2 \mu^* v_d  & b_\Psi & 0 & m_{\overline{\Psi}}^2+\Delta_{\bar{\Psi}}  \\
  \end{pmatrix}, \label{ms}
 \ee where $\Delta_\phi \equiv \frac{1}{2} (T_{3}g^2-Y_\phi g'^2)(v_d^2-v_u^2)$.
 
In the stop sector, the scalar masses are given by 
\begin{equation}
\frac{1}{2}\left(2m_t^2 + m_{\tilde{Q}}^2+m_{\tilde{u}}^2 + \Delta_{\tilde{t}_L} + \Delta_{\tilde{t}_R} \pm\sqrt{4m_t^2 X_t^2 +(m_{\tilde{Q}}^2+m_{\tilde{u}}^2 + \Delta_{\tilde{t}_L} - \Delta_{\tilde{t}_R})^2 }  \right), \label{stop_mass} 
\end{equation} where $X_t \equiv A_t - \mu \,{\rm cot} \beta$ and $m_{\tilde{Q}}^2,\,{\rm and}\,m_{\tilde{u}}^2$ are the $Q$ and $\overline{u}$ soft masses, which at the messenger scale are given by 
\begin{align}
m_{\tilde{Q}}^2 \,=& \left[ \frac{1}{8\pi^2}\left( \frac{4}{3}\alpha_3^2+\frac{3}{4}\alpha_2^2+\frac{1}{60}\alpha_1^2\right) \right. \\
&\left. + \frac{1}{256\pi^4}\left( 6y_t h_1 \kappa_{1,B}\lambda_{1,B} +y_t^2(9\kappa_{1,B}^2-2\kappa_{2,U}^2)+y_b^2(6\kappa_{2,B}^2-2\kappa_{2,U}^2) \right. \right. \nonumber \\
& \left. \left. +6\kappa_{1,B}^2(\kappa_{1,B}^2+\kappa_{2,U}^2+3\lambda_{1,B}^2)  -4\pi \kappa_{1,B}^2\left(\frac{8}{3}\alpha_3+\frac{3}{2}\alpha_2+\frac{13}{60}\alpha_1 \right) \right) \right] \left(\frac{F}{M} \right)^2, \nonumber\\
m_{\tilde{u}}^2 \,=& \left[ \frac{1}{8\pi^2}\left( \frac{4}{3}\alpha_3^2+\frac{4}{15}\alpha_1^2\right) + \frac{1}{256\pi^4}\left(  12 \kappa_{1,B}^4 + 3 \kappa_{2,U}^4 +4\kappa_{1,B}^2\kappa_{2,U}^2 \right. \right. \\
&\left. \left.  + 6 \kappa_{1,B}^2\lambda_{1,B}^2 +12 y_t h_1 \kappa_{1,B}\lambda_{1,B} +12 y_t^2 \kappa_{1,B}^2 +y_b^2(\kappa_{1,B}^2+\kappa_{2,U}^2) \right. \right. \nonumber \\
& \left. \left. +4\pi \kappa_{1,B}^2\left(160\alpha_1 + 90\alpha_2 + \frac{91}{30}\alpha_1\right)+4\pi \kappa_{2,U}^2\left(4\alpha_1 + \frac{89}{60}\alpha_1\right) \right) \right]\left(\frac{F}{M} \right)^2. \nonumber
\end{align} while the trilinear parameter is given by 
\begin{equation}
A_t\,=\,-\frac{1}{16\pi^2}\left(\frac{F}{M} \right)(3\kappa_{1,B}^2+3\kappa_{2,U}^2). 
\end{equation}

It is noteworthy the fact that the effect of having large values for $|A_t|$ on Equation (\ref{stop_mass}) is an increment in the splitting of the stop masses. However $A_t$ contributes directly to $m_{\tilde{Q}}^2$ and $m_{\tilde{u}}^2$, that implies that we can benefit from having relatively large $A$-terms as long as they are not much larger than the squark soft masses. A numerical depiction of $|A_t|$ can be seen in Figure \ref{Aterm} in the next section.
 
% 

%%%%%%%%%%%%%%%%%%%%%%%%%%%%%%%%%%%%%%%%%%%

%%%%%%%            HIGGS SECTION              %%%%%%%%%%%%

%%%%%%%%%%%%%%%%%%%%%%%%%%%%%%%%%%%%%%%%%%%

\section{The lightest CP-even Higgs mass}\label{higgsmass}

At tree level, the scalar potential for the Higgs fields is the same as in the MSSM \cite{Gunion:1989we,Martin:1997ns,Djouadi:2005gj}: 
\begin{align}
V^{\rm tree} = & \left( m_{H_u}^2+\mu^2\right) |H^0_u|^2 +\left( m_{H_d}^2+\mu^2\right) |H^0_d|^2 \label{tree-V}\\
& -b_\mu \left(H^0_u H^0_d + {\rm h.c.} \right) +\frac{1}{8}\left( g^2 + g'^2 \right) \left( |H^0_u|^2  - |H^0_d|^2 \right)^2. \nonumber
\end{align} In the mass spectrum, we find a charged pair of Higgs fields $H^\pm$, a CP-odd neutral scalar $A^0$ and two CP-even neutral scalars $h^0$ and $H^0$. As it is well known, at tree level, the mass of the lightest neutral Higgs is bounded from above by the mass of the $Z$ boson, .i.e. $m_{h} \le m_{Z}\, {\rm cos} 2\beta$. In the so-called ``decoupling limit", $m_A\gg m_Z$,  $m_h$ saturates this bound. This is the limit that we use in this work. 

In the MSSM scenario, at one-loop level, the top and stop fields contribute to the mass of the lightest Higgs. The effective potential due to these fields is 
\be 
V_{top}^{\rm 1-loop} =\frac{3}{32\pi^2}\sum_{i=1,2} {\rm Tr} \left[m_{\tilde{t}_i}^4 \left( {\rm Log} \frac{m_{\tilde{t}_i}^2}{Q^2}  -\frac{3}{2} \right) -m_{t_i}^4 \left( {\rm Log} \frac{m_{t_i}^2}{Q^2}  -\frac{3}{2} \right) \right]. 
\ee

The one-loop corrected mass of the lightest CP-even Higgs is, then, given by \cite{Haber:1996fp}
\be 
m_{h\,{\rm 1-loop}}^2 \approx m_Z^2 \,{\rm cos}^2 2\beta + \frac{3 g^2}{8 \pi^2}\frac{m_t^4}{m_W^2}\left[\, {\rm Log}\left(\frac{M_S^2}{m_t^2} \right)+ \frac{X_t^2}{M_S^2}\left(1 -\frac{X_t^2}{12\,M_S^2 } \right)\right], \label{mh-mssm}
\ee
with $M_S \equiv \sqrt{m_{\tilde{t}_1}m_{\tilde{t}_2}}$.  The two-loop correction to this expression is \cite{Carena:1995bx}
\begin{eqnarray}
\delta m^2_{h, {\rm 2-loop}} = &&\frac{3 g^2 m_t^4}{64 \pi^4 m_W^2} \left( \frac{3g^2m_t^2}{2m_W^2} -32 \pi \alpha_3 \right) \label{mh-2loop}\\ 
&&\times \left[ \frac{2X_t^2}{M_S^2}\left(1-\frac{X_t^2}{12 M_S^2}\right) +{\rm Log}\left(\frac{M_S^2}{m_t^2}\right)\right]{\rm Log}\left(\frac{M_S^2}{m_t^2}\right). \nonumber 
\end{eqnarray} For the numerical calculations in the following section, we will use both 1-loop and 2 -loop contributions to the Higgs mass, $m_{h\,{\rm MSSM}}^2 = m_{h\,{\rm 1-loop}}^2 + m^2_{h, {\rm 2-loop}}$.

Attaining a Higgs mass $m_h \approx 126 \,{\rm GeV}$ is somewhat challenging for minimal GMSB unless the top squarks have masses larger than $5 \,{\rm TeV}$, which, as already stated, generates some conflict with the naturalness of the MSSM. Additionally, in GMSB, the $A-$terms are zero at the messenger scale, although they are non-zero at the electroweak scale due to the running of their RGEs. However, in most minimal GMSB models, $X_t/M_S$ is still smaller than 1, which does not favor the enhancement of the Higgs mass coming from stop mixing; instead, the radiative corrections to $m_h$ are dominated by the logarithmic term in Equation (\ref{mh-mssm})\cite{Draper:2011aa,Ajaib:2012vc}. There are some scenarios in GMSB where stops lighter than $5 \,{\rm TeV}$ and $X_t/M_S >1$ are possible, by having heavy gauginos or a very large messenger scale \cite{Draper:2011aa}.

When we introduce the new vector-like fields, there is a similar effective one-loop potential from the vector-like fields,
\be 
V_{VL}^{\rm 1-loop} =\frac{3}{32\pi^2} \sum_{i} {\rm Tr}\left[m_{S_i}^4 \left( {\rm Log} \frac{m_{S_i}^2}{Q^2}  -\frac{3}{2} \right) -m_{F_i}^4 \left( {\rm Log} \frac{m_{F_i}^2}{Q^2}  -\frac{3}{2} \right) \right], 
\label{1loop-V}
\ee where $i$ runs over $(\Phi, \Psi, \bar{\Phi}, \bar{\Psi})$ and $m_{F_i}^2,\, m_{S_i}^2$ are the eigenvalues of the mass matrices (\ref{mf}, \ref{ms}) respectively. The correction to the Higgs mass is given by \cite{Martin:2009bg}

\begin{eqnarray} 
\delta m_h^2 =&& \left[\frac{{\rm sin}^2\beta}{2}\left( \frac{\partial^2 }{\partial v_u^2}-\frac{1}{v_u}\frac{\partial }{\partial v_u}\right) + \frac{{\rm cos}^2\beta}{2}\left( \frac{\partial^2 }{\partial v_d^2}-\frac{1}{v_d}\frac{\partial }{\partial v_d}\right) +{\rm sin}\beta\,{\rm cos} \beta \frac{\partial^2}{\partial v_u \partial v_d}\right] V_{VL}^{\rm 1loop}. \nonumber \\
\approx && \frac{3}{4\pi^2}\frac{m_W^2}{g^2}  h_1^4 \sin^2\beta \left[ {\rm Log} \left( \frac{M_{S,vector}^2+M_{10}^2}{m_{10}^2} \right) \right.  \label{deltamhsq}  \\ 
&&\left. +\frac{X_1^2}{12}\left( \frac{4(3M_{S,vector}^2+2M_{10}^2)-X_1^2-8 M_{S,vector}^2 M_{10}^2-10 M_{S,vector}^2}{(M_{S,vector}^2+M_{10}^2)^2}\right)\right], \nonumber
\end{eqnarray} where $M_{S, {\rm vector}}\equiv \sqrt{m_{\tilde{\Phi}} m_{\tilde{\Psi}}}$ and $X_1 \equiv A_1 - \mu \,{\rm cot} \beta$.

The correction to the Higgs mass is, then, given by 
\be 
\Delta m_h = \sqrt{m^2_{h,\,{\rm MSSM}} + \delta m_h^2} \,-\,m_{h,\,{\rm MSSM}} . \label{deltamh}
\ee

\subsection{Parameter scan}

In order to analyze the implications of this correction to the Higgs mass, we numerically scan the parameter space over the ranges shown in Table \ref{par_range}. We use these input parameters to compute the soft terms at the messenger scale and, then, we use the one-loop RG equations to extract the value of the different masses and couplings at the electroweak scale. For each set of input parameters we calculate the Higgs mass with and without the extra vector-like fields. 

In our parameter scan, we vary $h_1$ and $h_2$ such that the model is perturbative below the unification scale \footnote{In this work, the unification scale is defined as the energy scale where $g_1$ and $g_2$ meet.}. $h_2$ does not play an important role in raising $m_h$, hence we assume $h_2\ll h_1$. The best results for $\Delta m_h$ are obtained for $h_1\approx 1.0$. 

\begin{figure}[ht]\centering
\bigskip
\includegraphics[width=3.2in]{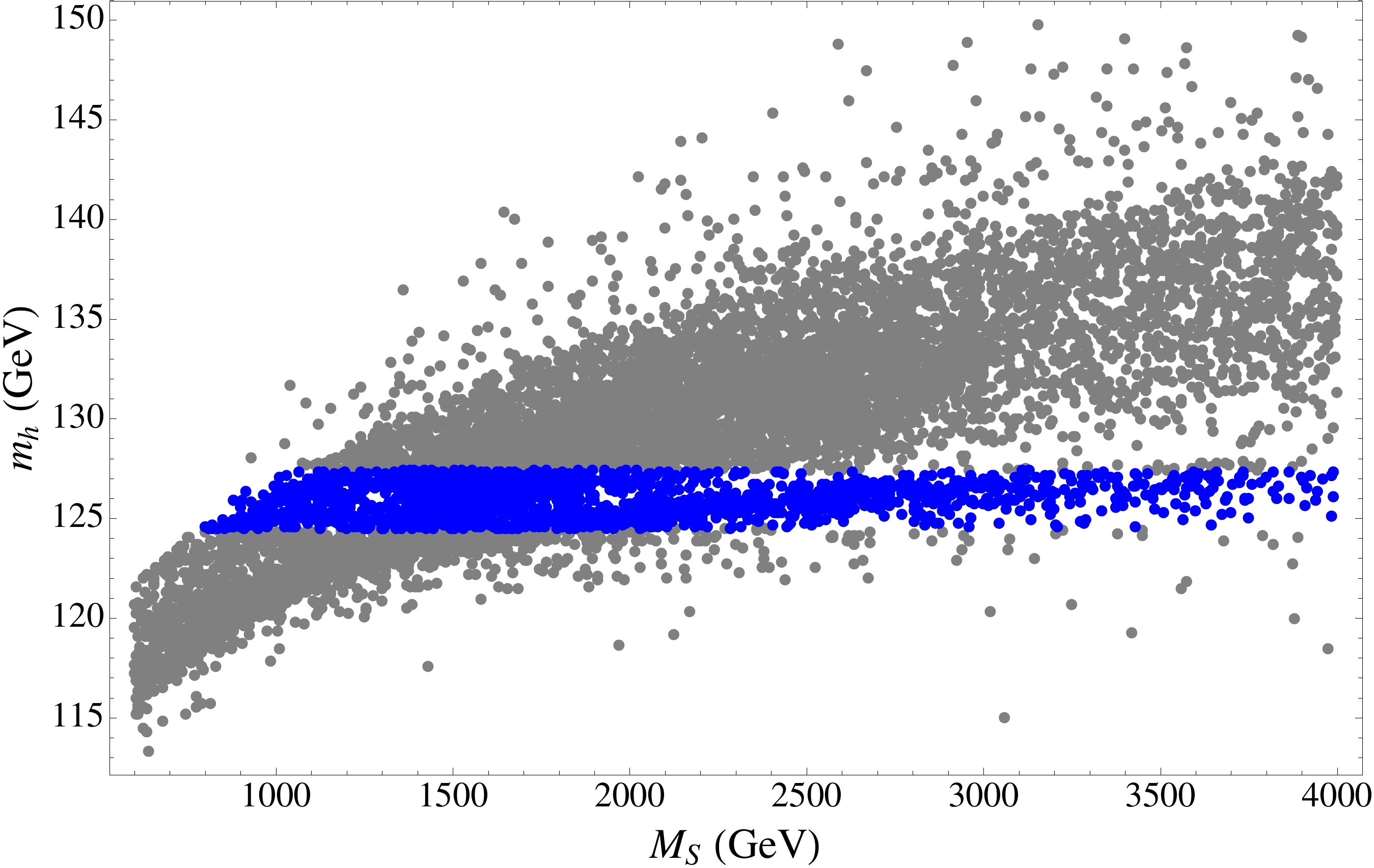} ~~~
\caption{$m_{h}$ values for different $M_S$. The blue dots denote the area around 126 GeV. Notice that, even for stops lighter than 4 TeV,  the addition of the new fields can lift the mass of the Higgs well above the values achieved in the minimal GMSB scenario.} \label{fig:mhMS}
\bigskip
\end{figure}

For simplicity, we have set the mass of the vector-like fermions to be the same at the GUT scale. At the electroweak scale, this fermions have vector-like masses between 700 GeV and 2 TeV. Figure \ref{fig:mhMS} shows the values of $m_h$ vs $M_S$ when the vector-like fields are included. It is important to note that $m_h\approx 126$ GeV is easily achieved in this model even for stop masses under 1 TeV. In the analyzed parameter space, the correction to the Higgs mass, given in equation (\ref{deltamh}), can be as small as $1\,{\rm GeV}$ and and as large as 35 GeV. However, when we just look at the region where the corrected mass of the Higgs is $125.9\pm 2.0 \,{\rm GeV}$ (the blue dots in Figure \ref{fig:mhMS}), we find that $\Delta m_h$ takes values between $1 \,{\rm GeV}$ and $10 \,{\rm GeV}$. This enhancement is more notable for vector-like masses around $1\,{\rm TeV}$.

\begin{figure}[ht]\centering
\bigskip
\includegraphics[width=3.2in]{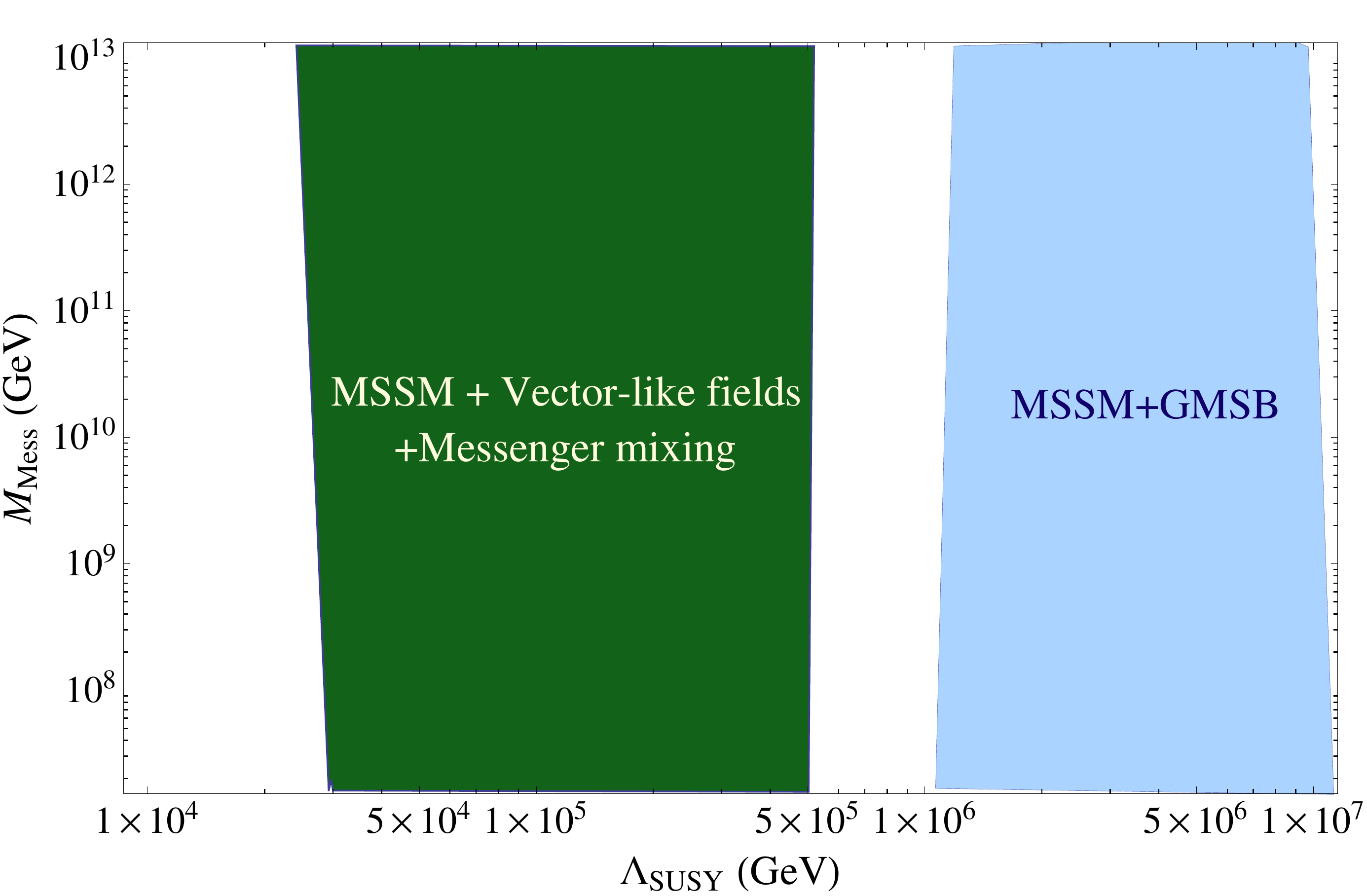} ~~~
\caption{Regions of the $M_{\rm mess}$ vs $\Lambda_{SUSY}$ plane that are consistent with $m_h\approx 126$ GeV.} \label{MmessLambda}
\bigskip
\end{figure}

As for the SUSY scale $\Lambda_{SUSY}$ that corresponds to $m_h\approx 126 \,{\rm GeV}$ and $m_{\tilde{t}_1}<2 $ TeV, we obtained values as low as $3\times 10^4$ GeV and as high as $6\times 10^5$ GeV with messengers masses between $10^{7}$ and $10^{13}$ GeV.  Notice that this implies a lower scaler for $\Lambda_{SUSY}$ than in the GMSB scenario, where ${\rm min} [\Lambda_{SUSY}]\approx 5\times 10^5$ GeV \cite{Ajaib:2012vc}. A comparison of these scales is depicted in Figure \ref{MmessLambda}.

\begin{table}[h]
\centering
\begin{tabular}{|c|c|}
\hline 
Parameter & Range \\ 
\hline
$\Lambda_{SUSY} \equiv F/M$ & $[10^4,\, 10^6] \,{\rm GeV}$ \\
\hline

$M$ &$ [10^7, \,10^{13}] \,{\rm GeV}$ \\
\hline
$M_{10}$ &$ [300, \,1400] \,{\rm GeV}$ \\
\hline 
${\rm Tan}\,\beta$ &$[10,\,50]$ \\
\hline
$h_1$ $(h_2 < h_1)$ &$[0.5,\,1.2]$ \\
\hline
$\lambda_{i},\,\kappa_{j}$  &$[0.05,\,0.3]$ \\ 
\hline
\end{tabular}
\label{par_range}
\caption{Input parameters and their ranges used in the numerical calculations. The values for $h_1$ and $h_2$ are given at the electroweak scale, whereas the values for $\lambda_i$ and $\kappa_i$ are given at the messenger scale.}
\end{table}

\begin{figure}[ht]\centering
\bigskip
\includegraphics[width=3.2in]{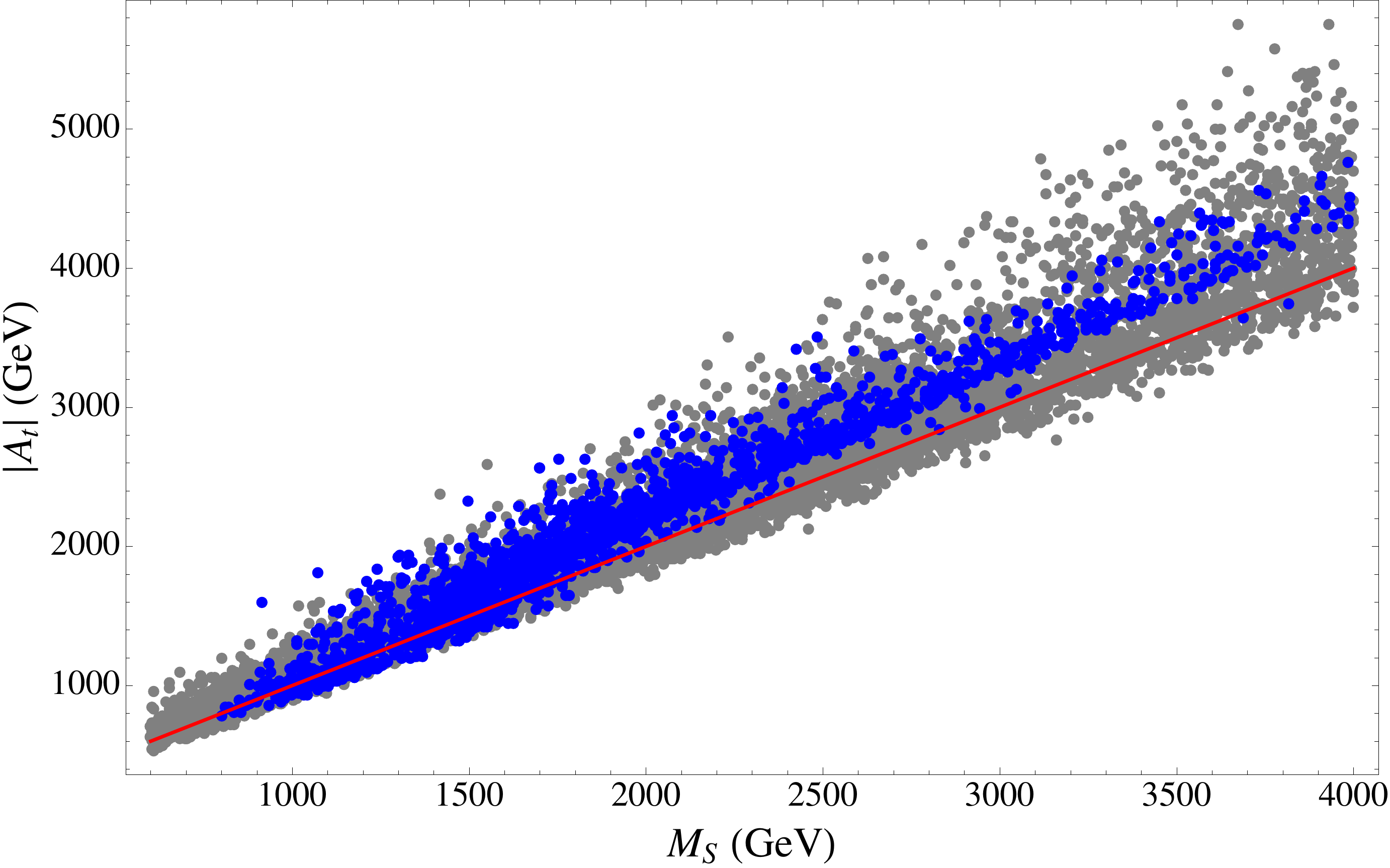} ~~~
\caption{Plot of $|A_t|$ vs $M_S$. The blue dots correspond to the cases where $m_h\approx 126$ GeV. The red line depicts $|A_t|=M_S$.} \label{Aterm}
\bigskip
\end{figure}

%\begin{figure}[ht]\centering
%\bigskip
%\includegraphics[width=3.in]{dmMphi_all.eps} ~~~
%\caption{Improvement to $m_{h,\,{\rm MSSM}}$ obtained  in Equation (\ref{deltamh}). The dark shaded region corresponds to the case when $m_h =\sqrt{m^2_{h,\,{\rm MSSM}} + \delta m_h^2}\approx 125.9\pm0.5 \,{\rm GeV}$.  The light gray area shows the corrections obtained in those cases where $m_h$ is larger or smaller than the measured Higgs mass.} \label{fig:dmMphi}
%\bigskip
%\end{figure}
%
Now, we take a look at the values of the $A-$terms when we include the vector-like fields. Figure \ref{Aterm} shows the values for $A_t$ for different values of $M_S$. It can be seen that, indeed, it is possible to obtain trilinear terms such that $|A_t/M_S|\,>1$, unlike the MSSM case with minimal GMSB. Thus, this feature of the model aides the addition of the vector-like fields in the lifting of the Higgs mass. It is worth quantifying the effect of the messenger-matter mixing to the values of $A_t$ compared to the mGMSB scenario. Defining $\delta A\equiv A_t - A_{t,\,{\rm no\, mixing}}$ at the low scale, we find that 
\begin{equation}
\abs{\frac{\delta A}{A_t}}_{M \sim 10^9 \,{\rm GeV}} \sim 0.25,\qquad \abs{\frac{\delta A}{A_t}}_{M \sim 10^{10} \,{\rm GeV}} \sim 0.20 , \qquad \abs{\frac{\delta A}{A_t}}_{M \sim 10^{11} \,{\rm GeV}} \sim 0.18.
\end{equation}

\begin{figure}[ht]\centering
\bigskip
\includegraphics[width=3.2in]{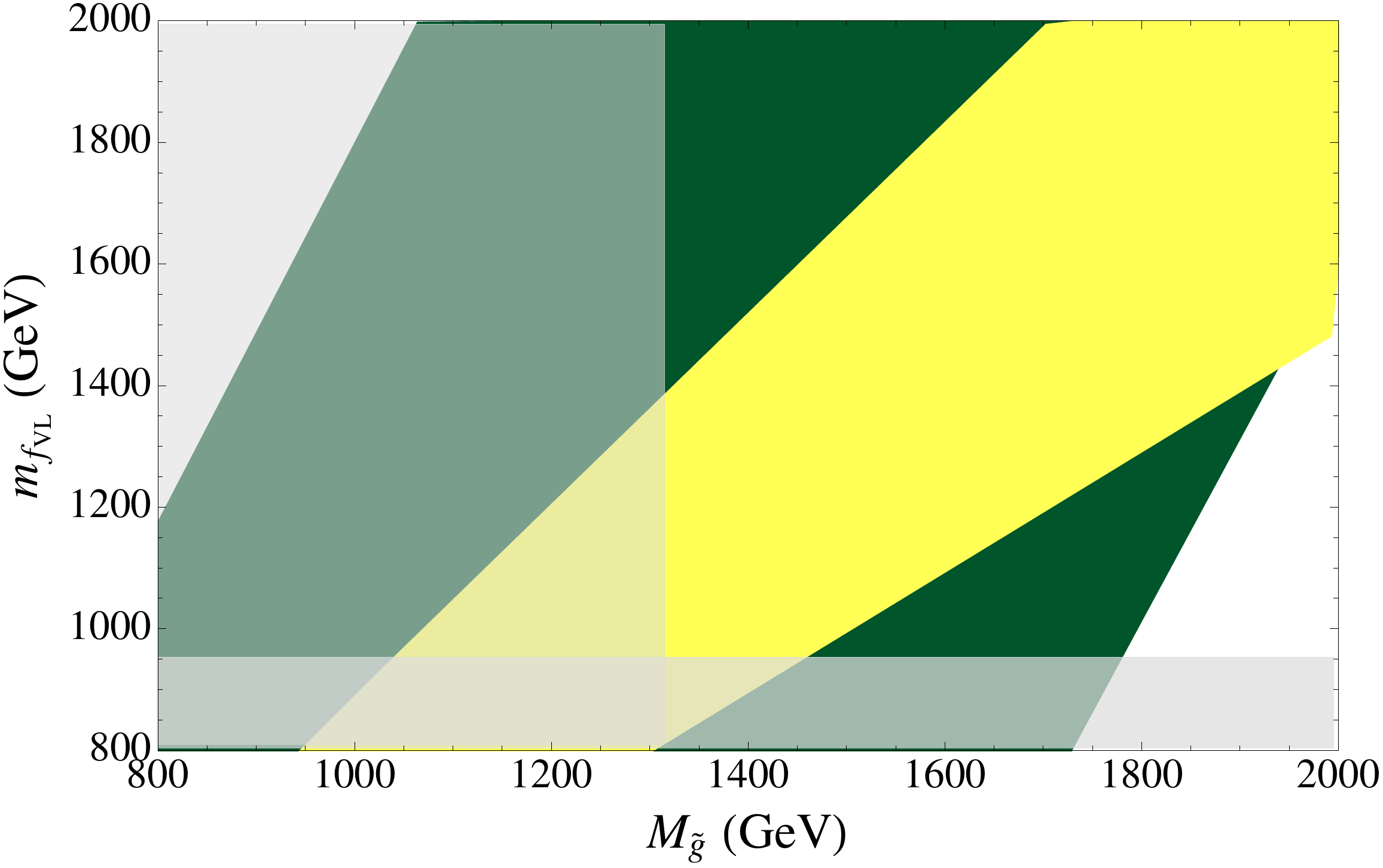} ~~~
\caption{Mass of the vector-like fermions vs gaugino mass for models with (green) and without (yellow) messenger mixing. The gray shaded areas correspond to the estimated exclusions by lower bounds on gluino and vector-like fermion masses (see next section).} \label{MG}
\bigskip
\end{figure}

To gain further insight about how having larger $A$ terms helps in raising $m_h$, we compare, in Figure \ref{MG}, the regions in the vector-like fermion mass $M_{VL}$ vs $M_{\tilde{G}}$ plane that are compatible with a Higgs mass of $126 \pm 2$ GeV for two cases: our model with messenger mixing (green area) and a model without mixing (yellow area, see \cite{Martin:2012dg}). This is shown for $h_1=1$, ${\rm tan}\beta =20$, $M=10^{10}$ GeV and $1.6>|A_t/M_S|>1.4$ . It can be noticed that, even if the yellow region shows that a realistic mass for the Higgs can be obtained with low masses for the vector-like fermions and the gaugino, the addition of the couplings to the messengers improves the possibilities.

Considering the fact that the Higgs mass is easily lifted in this type of models and the mass of the stops is not required to be larger than 2 TeV, one could think naively that such a proposal is more ``natural" than the MSSM with mGMSB. However, as pointed out in \cite{Martin:2009bg}, the new vector-like fields will contribute to the soft term $-m_{H_u}^2$, in the same way than stops,  through the radiative term
\be 
\Delta m_{H_u}^2 \approx -\frac{3 h_1^2}{4\pi^2}M_{S, {\rm vector}}^2 \,{\rm ln} \frac{M}{M_{S, {\rm vector}}},
\ee where $M_{S, {\rm vector}}\equiv \sqrt{m_\Phi m_\Psi}$.  This, apparently, would make the little hierarchy problem even worse. However, the effect of this term, and the stop contribution, in making $-m_{H_u}^2$ large is smaller in this case (where the stops and the vector sfermions have masses just above 1 TeV) compared to the case where 5 TeV stops are required. Concretely,
\be 
\frac{(m_{H_u}^2)_{\rm This\, model}}{(m_{H_u}^2)_{\rm GMSB}} \sim 0.1.
\ee
 Thus, this effect together with the correction to the Higgs mass lead us to argue that there is some alleviation of the fine tuning problem existing in the MSSM with mGMSB. In fact, if we consider the measure \cite{Barbieri:1987fn}
\begin{equation}
\Delta \equiv Max\left[\abs{\frac{\partial\, {\rm Log} M_Z^2}{\partial\, {\rm Log} a_i^2}}\right],  
\end{equation} where $a_i \in (h {\rm's},\, \lambda {\rm's},\,M_{10}, M,\, \Lambda_{SUSY})$, this model produces values for $\Delta$ between 300 and 500 for the range of parameters studied in this section and corresponding to the green region in Figure \ref{MmessLambda}. This turns out to be an order of magnitude smaller than the values of $\Delta$ corresponding to mGMSB in the blue region of the same figure. It is interesting to remark that for one of the least tuned points in the parameter space, which is depicted in Figure \ref{spectrum}, the contribution to the Higgs mass enhancement due to the presence of the vector-like fields is dominant respect to the fact of having a large $A_t$ term. 

\subsection{Some phenomenology comments}

\begin{figure}[ht]\centering
\bigskip
\includegraphics[width=3.5in]{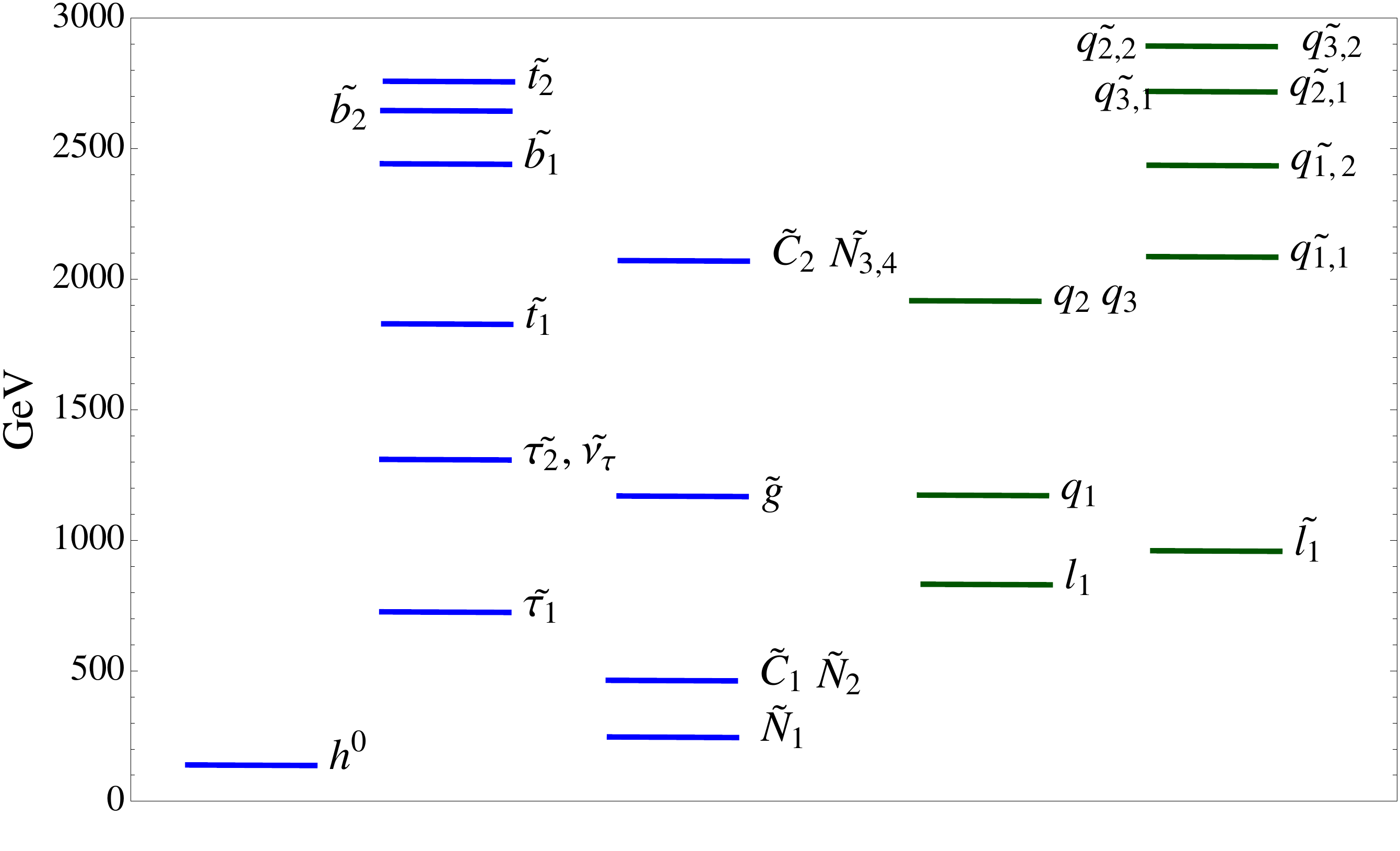} ~~~
\caption{Example of a low energy spectrum for $M=9\times 10^8\,{\rm GeV}$, $\Lambda = 1.4 \times 10^5\,{\rm GeV}$ and $\tan \beta = 15.6$} \label{spectrum}
\bigskip
\end{figure}

For a more general discussion about the LHC signatures of this type of models with extra vector-like matter, we refer the reader to \cite{Martin:2012dg}. For the sake of clarity, we present a sample spectrum in Figure \ref{spectrum}, where the new charged quarks (combinations of the $\Phi$ and $\Psi$ fermions) are denoted as $q_{i}$ and their spartners are $\tilde{q}_{i,j}$ with $j=1,2$. Likewise, the charged vectorlike lepton is denoted as $l$ and its spartners $l_i$. 

In our model, the lightest neutralino has a mass $M_{\tilde{N}_1} \gtrsim 140$ GeV and $m_{\tilde{\tau_1}} \gtrsim 600$ GeV, in which case the limits set by the LHC searches using simplified models and CMSSM/mSUGRA are applicable to our analysis \cite{Aad:2012fqa, Chatrchyan:2013oca}. This implies a lower bound on the gluino mass around 1.3 TeV  (see Figure \ref{MG}) and 500 GeV for the stau mass\footnote{ The discovery possibilities come mainly from the processes $pp\rightarrow \,\tilde{g}\tilde{g},\, \tilde{g}\tilde{Q},\,\tilde{C}_1^+\tilde{C}_1^-$. Some of the gluinos decay to jets and charginos and, then, the wino-charginos decay through the channels $\tilde{C}_1\rightarrow \,\tilde{\tau}_1\,\nu\rightarrow\,\tau\, \tilde{N}_1\,\nu$.}. 

Since the MSSM matter does not mix, at tree level, with the vector-like fermions, the lightest of these, $q_1$ (a combination of $\Phi$ and $\Psi$ fermions), is long-lived. For $m_{q_1}\sim 1.3$ TeV, its lifetime is $10^{9}\,{\rm s} \gtrsim \tau \gtrsim 10^{5}$ s. There exist cosmological bounds on this type of relics, coming from photodissociation of light elements\cite{Kawasaki:1994sc}. After computing the number density of these colored particles \cite{Kang:2006yd,Berger:2008ti}, we find that $n_{q_1}/s \sim 10^{-18}$, where $s$ is the entropy density and $n_{q_1}$ is the relic number density. This is consistent with the cosmological constraints. 

On the collider side, searches for a charged massive stable particle have been done at the LHC at collision energies of $\sqrt{s} = 7$ and $8$ TeV\cite{Aad:2012pra, Chatrchyan:2013oca}, where these particles can be identified by their anomalous energy loss or a long time-of-flight to the outer detectors. The limit obtained in \cite{Martin:2012dg} for the mass of this fermion is $m_{q_1} \gtrsim 950$ GeV, which constraints the bottom part of Figure \ref{MG}, but still leaves a large region in the parameter space where the vector-like fermion mass is above 1 TeV.

Finally, for $\Lambda_{SUSY} \sim 10^5$ GeV, cosmological gravitino constraints set some bounds on the mass of the messengers and the reheating temperature after inflation. In order to avoid overclosure of the universe by the thermal relic abundance of the gravitino, the reheat temperature should be lower than $10^4$ GeV. The next constraint comes from big bang nucleosynthesis (BBN), which sets a bound on the gravitino mass such that the NLSP decays to a photon and a gravitino before $t_{\rm BBN} \approx 0.1 \,{\rm s}$ (see \cite{Kohri:2005wn} and references therein). This implies $m_{3/2} \lesssim 1$ MeV, which translates into $M\lesssim 2\times 10^{11}$ GeV.

\section{Conclusions}

We have presented a model that yields a Higgs mass around $126\,{\rm GeV}$ with stop masses under 2 TeV. In order to do so, we have synthesized two approaches used in the literature to tackle the little hierarchy problem. First, we have added a new set of fields to the effective theory. These fields possess vector-like masses around 1 TeV and contribute to raise the mass of the Higgs field through radiative corrections. On the other hand, we have included in the superpotential several marginal terms that couple the low energy degrees of freedom to the messenger sector. This provides a mechanism to enhance the trilinear $A$ terms in the soft Lagrangian. This effect complements the enhancement of the Higgs mass by increasing the one-loop corrections through the stop (and vector-like scalar) mixing. This way, we obtained a wider region in the parameter space for which the measured mass of the Higgs can be achieved, while keeping the sfermions with relatively small masses. In addition, the soft term $m_{H_u}^2$ is smaller for this type of models. These facts allow us to argue that the fine tuning problem in the MSSM is  improved by this construction.

So far, no signature of new vector matter or SUSY has been observed at the LHC. However, since this sort of models contains stops and vector-like sfermions with masses just above the TeV scale, it presents a spectrum that can be searched for in the near future.

\acknowledgments 

We are grateful to Michael Dine for providing very useful comments and for reminding W.F. about their work done a few decades ago. W.T. also thanks Jianghao Yu, Can Kilic, Alejandro de la Puente and Simon Knapen for insightful discussions. We also thank Renato Fonseca for pointing out a couple of issues in the original version of this article. This material is based upon work supported by the National Science Foundation under Grant Numbers PHY-1316033 and PHY-0969020 and by the Texas Cosmology Center.
\appendix
\section{Soft terms}\label{appendix1}

We calculate the soft masses at the messenger scale based on \cite{Evans:2013kxa}, using the superpotential \ref{SuperW}. In such work, the formulas for the soft parameters are derived from wave function renormalization taking into account the discontinuity of the wave functions at the messenger threshold. 

In addition to the soft mass contribution from the standard gauge mediation
\be 
m^2_{\phi_i}= 2\left( \frac{F}{M}\right)^2 \left[ \left( \frac{\alpha_3}{4\pi}\right)^2 C_3(i) +\left(\frac{\alpha_2}{4\pi}\right)^2 C_2(i) +\frac{3}{5}\left(\frac{\alpha_1}{4\pi}\right)^2 Y_i^2 \right], 
\ee there is a two-loop contribution due to the Yukawa couplings between the messenger sector and the effective theory. 
\begin{align}
m^2_{\phi_a,\, {\rm Yuk}} = &\frac{1}{256\pi}\left( \frac{F}{M}\right)^2 \left[ \frac{1}{2}d^{jk}_a d^{lm}_i \left (\Delta(\lambda^*_{aik}\lambda_{ajk} )(\lambda_{ilm}\lambda^*_{jlm})^+-(\lambda^*_{aik}\lambda_{ajk})^- \Delta(\lambda_{ilm}\lambda^*_{jlm} \right) \right. \\
& \left. +\frac{1}{4} d_a^{ij}d_a^{kl} \Delta(\lambda^*_{aij}\lambda_{cij}) \Delta(\lambda^*_{ckl}\lambda_{bkl})- d_a^{ij} C_r^{aij} g_r^2  \Delta(\lambda^*_{aij}\lambda_{aij}) \right],\nonumber
\end{align} where $d_i^{jk}$ is a multiplicity factor, $C_r^{aij} = C_r^a+C_r^i+C_r^j$ sums the quadratic Casimirs of the fields, and the sum over $c$ includes only the MSSM matter.  There is also a subleading one-loop contribution
\be 
m^2_{\phi_a, {\rm 1-loop}} \approx \frac{1}{24\pi} \left( \frac{F^2}{M^3}\right)^2 d_a^{ij}  \Delta(\lambda^*_{aij}\lambda_{aij}).
\ee

On the other hand, the $A$-terms are given by
\be 
A_{\phi_a} \approx \frac{1}{32\pi^2} \left( \frac{F}{M}\right)  d_a^{ij}  \Delta(\lambda^*_{aij}\lambda_{aij}).
\ee
%\begin{align}
%m^2_{Q,\, {\rm Yuk}} =& \frac{1}{256\pi^4} \left(\frac{F }{M} \right)^2 \left[ \kappa_{1,A}^2(3\kappa_{1,B}^2+3\kappa_{2,L}^2 +3\kappa_{2,Q}^2 +y_t^2+y_b^2) + 3\kappa_{1,B}^4+6\kappa_{1,A}^4  \right.\\ 
%  & \left. +  \kappa_{1,B}^2(3\lambda_{1,B}^2+2\kappa_{2,L}^2 +12\kappa_{2,Q}^2 + 9 y_t^2+\kappa_{2,u}^2)+6y_t\, h_1 \kappa_{1,B}\,\lambda_{1,B}   \right. \nonumber \\
% & \left. +6\lambda_{3,\Phi}^4+\lambda_{3,\Phi}^2\lambda_{3,\chi}^2+(4\lambda_{3,\Phi}^2-2h_1^2)\lambda_{3,\Psi}^2 - \frac{4\pi}{3}\alpha_1(3\lambda_{1,A}^2 +13 \lambda_{1,B}^2-56 \lambda_{3,\Phi}^2)  \right. \nonumber \\
%  & \left.   -6\pi \alpha_{2} (\lambda_{1,A}^2+\lambda_{1,B}^2+8\lambda_{3,\Phi}^2)-\frac{16\pi}{3}\alpha_{3}(3\lambda_{1,A}^2+2(\lambda_{1,B}^2+8\lambda_{3,\Phi}^2)) \right]. \nonumber 
%\end{align}
 
\section{Renormalization group equations}\label{appendix2}

\begin{align}
\frac{d g_i}{dt}\,&=\,\frac{b_i}{16\pi^2} g_i^3,\qquad b_i\,=\,(33/5 + 3,\, 3+1, \,-3+3),\quad t\equiv {\rm log}\left( \frac{Q}{\rm GeV}\right).\\
\frac{d M_i}{dt}\,&=\,\frac{b_i}{8\pi^2} g_i^2\,M_i.\nonumber
\end{align}

\begin{align}
16 \pi^2 \frac{d y_t}{dt} &= y_t\left(6y_t^2+y_b^2+3 h_1^2-\frac{16}{3}g_3^2-3g_2^2-\frac{13}{15}g_1^2\right), \\
16 \pi^2 \frac{d y_b}{dt} &= y_b\left(y_t^2+6y_b^2+y_\tau^2 +3 h_2^2-\frac{16}{3}g_3^2-3g_2^2-\frac{7}{15}g_1^2\right), \nonumber \\
16 \pi^2 \frac{d y_\tau}{dt} &= y_\tau\left(3y_b^2+4 y_\tau^2 +3 h_2^2-3g_2^2-\frac{9}{5}g_1^2\right), \nonumber \\
16 \pi^2 \frac{d h_1}{dt} &= h_1\left(3y_t^2+ h_1^2-\frac{16}{3}g_3^2-3g_2^2-\frac{13}{15}g_1^2\right), \nonumber\\
16 \pi^2 \frac{d h_2}{dt} &= h_2\left(3y_b^2+y_\tau^2+ h_2^2-\frac{16}{3}g_3^2-3g_2^2-\frac{13}{15}g_1^2\right). \nonumber
\end{align}

\begin{align}
16 \pi^2 \frac{d a_t}{dt} &= a_t\left(18 y_t^2+y_b^2+9 h_1^2-\frac{16}{3}g_3^2-3g_2^2-\frac{13}{15}g_1^2\right) \\
&\quad +2 a_b y_b^* y_t +2 a_2 h_2^* y_t + y_t \left(16 g_3^2 M_3 +6 g_2^2 M_2 +\frac{26}{15}g_1^2 M_1 \right), \nonumber \\
16 \pi^2 \frac{d a_b}{dt} &= a_b\left(y_t^2+18y_b^2+y_\tau^2 +9 h_2^2-\frac{16}{3}g_3^2-3g_2^2-\frac{7}{15}g_1^2\right) \nonumber \\
&\quad +2 a_t y_b^* y_t +2 a_\tau y_b^* y_\tau ++2a_1 h_1^* y_b+ y_b \left(16 g_3^2 M_3 +6 g_2^2 M_2 +\frac{14}{15}g_1^2 M_1 \right), \nonumber \\
16 \pi^2 \frac{d a_\tau}{dt} &= a_\tau\left(3y_b^2+12y_\tau^2 +3 h_2^2-3g_2^2-\frac{9}{5}g_1^2\right) \nonumber \\
&\quad +6 a_b y_b^* y_\tau +6 a_2 h_2^* y_\tau + y_\tau \left(6 g_2^2 M_2 +\frac{18}{5}g_1^2 M_1 \right), \nonumber \\
16 \pi^2 \frac{d a_1}{dt} &= a_1\left(3y_t^2+ 18 h_1^2-\frac{16}{3}g_3^2-3g_2^2-\frac{13}{15}g_1^2\right) \nonumber\\
&\quad +2 a_2 h_2^* h_1 +2 a_t y_t^* h_1 + h_1 \left(16 g_3^2 M_3 +6 g_2^2 M_2 +\frac{26}{15}g_1^2 M_1 \right), \nonumber \\
16 \pi^2 \frac{d a_2}{dt} &= a_2\left(3y_b^2+y_\tau^2+ 18h_2^2-\frac{16}{3}g_3^2-3g_2^2-\frac{13}{15}g_1^2\right) \nonumber\\
&\quad +2 a_1 h_1^* h_2 +2 a_b y_b^* h_2 + 2a_\tau y_\tau^* h_2+ h_2 \left(16 g_3^2 M_3 +6 g_2^2 M_2 +\frac{26}{15}g_1^2 M_1 \right). \nonumber 
\end{align}

\begin{align}
X_t(t)&\equiv 2\abs{y_t}^2(m_{H_u}^2+m_{\tilde{Q}}^2+m_{\tilde{u}}^2)+2\abs{a_t}^2\\
X_b(t)&\equiv 2\abs{y_b}^2(m_{H_d}^2+m_{\tilde{Q}}^2+m_{\tilde{d}}^2)+2\abs{a_d}^2 \nonumber \\
X_\tau(t)&\equiv 2\abs{y_\tau}^2(m_{H_d}^2+m_{\tilde{L}}^2+m_{\tilde{e}}^2)+2\abs{a_\tau}^2 \nonumber \\
X_1(t)&\equiv 2\abs{h_1}^2(m_{H_u}^2+m_{\tilde{\Psi}}^2+m_{\tilde{\Phi}}^2)+2\abs{a_1}^2  \nonumber \\
X_2(t)&\equiv 2\abs{h_2}^2(m_{H_d}^2+m_{\tilde{\bar{\Psi}}}^2+m_{\tilde{\bar{\Phi}}}^2)+2\abs{a_2}^2  \nonumber \\
S(t)&\equiv m_{H_u}^2 -m_{H_d}^2+m_{\tilde{Q}}^2 +m_{\tilde{d}}^2 +m_{\tilde{e}}^2 -m_{\tilde{L}}^2-2m_{\tilde{u}}^2  \nonumber \\
&+m_{\tilde{\Phi}}^2 -m_{\tilde{\bar{\Phi}}}^2 + m_{\tilde{\Phi}}^2 +m_{\tilde{\bar{\Phi}}}^2 - 2 m_{\tilde{\Psi}}^2 + 2m_{\tilde{\bar{\Psi}}}^2 + m_{\tilde{\chi}}^2 -m_{\tilde{\bar{\chi}}}^2 \nonumber
\end{align}

\begin{align}
16 \pi^2 \frac{d m_{\tilde{Q}}^2}{dt} &= X_t + X_b -\frac{32}{3} g_3^2 M_3^2 - 6 g_2^2 M_2^2 - \frac{2}{15} g_1^2 M_1^2 +\frac{1}{5}S \\
16 \pi^2 \frac{dm_{\tilde{u}}^2}{dt} &=2 X_t  -\frac{32}{3} g_3^2 M_3^2 - \frac{32}{15} g_1^2 M_1^2 -\frac{4}{5}S \nonumber \\
16 \pi^2 \frac{dm_{\tilde{d}}^2}{dt} &= 2 X_b  -\frac{32}{3} g_3^2 M_3^2 - \frac{8}{15} g_1^2 M_1^2 +\frac{2}{5}S \nonumber \\
16 \pi^2 \frac{d m_{\tilde{L}}^2}{dt} &= X_\tau - 6 g_2^2 M_2^2 - \frac{6}{5} g_1^2 M_1^2 -\frac{3}{5}S \nonumber \\
16 \pi^2 \frac{dm_{\tilde{e}}^2}{dt} &= 2 X_\tau  - \frac{24}{5} g_1^2 M_1^2 +\frac{6}{5}S \nonumber \\
16 \pi^2 \frac{d m_{H_u}^2}{dt} &= 3X_t+X_1 - 6 g_2^2 M_2^2 - \frac{6}{5} g_1^2 M_1^2 +\frac{3}{5}S \nonumber \\
16 \pi^2 \frac{d m_{H_d}^2}{dt} &=3 X_b +X_2- 6 g_2^2 M_2^2 - \frac{6}{5} g_1^2 M_1^2 -\frac{3}{5}S \nonumber \\
16 \pi^2 \frac{d m_{\tilde{\Phi}}^2}{dt} &= X_1  -\frac{32}{3} g_3^2 M_3^2 - 6 g_2^2 M_2^2 - \frac{2}{15} g_1^2 M_1^2 +\frac{1}{5}S \nonumber \\
16 \pi^2 \frac{dm_{\tilde{\Psi}}^2}{dt} &=2 X_1  -\frac{32}{3} g_3^2 M_3^2 - \frac{32}{15} g_1^2 M_1^2 -\frac{4}{5}S \nonumber \\
16 \pi^2 \frac{d m_{\tilde{\bar{\Phi}}}^2}{dt} &= X_2 -\frac{32}{3} g_3^2 M_3^2 - 6 g_2^2 M_2^2 - \frac{2}{15} g_1^2 M_1^2 -\frac{1}{5}S \nonumber \\
16 \pi^2 \frac{dm_{\tilde{\bar{\Psi}}}^2}{dt} &=2 X_2  -\frac{32}{3} g_3^2 M_3^2 - \frac{32}{15} g_1^2 M_1^2 +\frac{4}{5}S \nonumber \\
16 \pi^2 \frac{dm_{\tilde{\chi}}^2}{dt} &= - \frac{24}{15} g_1^2 M_1^2 +\frac{1}{5}S \nonumber \\
16 \pi^2 \frac{dm_{\tilde{\bar{\chi}}}^2}{dt} &= - \frac{24}{15} g_1^2 M_1^2 -\frac{1}{5}S \nonumber 
\end{align}

\end{document}